# Lab-on-a-Chip Optical Biosensor Platform: Micro Ring Resonator Integrated with Near-Infrared Fourier Transform Spectrometer

Kyoung Min Yoo[1,] *, May Hlaing[2], Sourabh Jain[1], James Fan[1], Yue An[1], and Ray T. Chen[1, 2,] *

[1]Department of Electrical and Computer Engineering, The University of Texas at Austin, 10100 Burnet Rd. Austin, TX, 78758, USA.

[2]Omega Optics Inc., 8500 Shoal Creek Blvd., Bldg. 4, Suite 200, Austin, TX, 78757, USA.

*Corresponding author. E-mail: chenrt@austin.utexas.edu, yoo_eb@utexas.edu

# ABSTRACT

A micro-ring-resonator (MRR) optical biosensor based on the evanescent field sensing mechanism has been extensively studied due to its high sensitivity and compact device size. However, a suitable on-chip integrated spectrometer device has to be demonstrated for the lab-on-a-chip applications, which can read the resonance wavelength shift from MRR biosensors based on minuscule changes in refractive index. In this paper, we demonstrated the design and experimental results of the near-infrared lab-on-a-chip optical biosensor platform that monolithically integrates the MRR and the on-chip spectrometer on the silicon-on-insulator (SOI) wafer, which can eliminate the external optical spectrum analyzer for scanning the wavelength spectrum. The symmetric add-drop MRR biosensor is designed to have a free spectral range (FSR) of ~19 nm, and a bulk sensitivity of ~73 nm/RIU; then the drop-port output resonance peaks are reconstructed from the integrated spatial-heterodyne Fourier transform spectrometer (SHFTS) with the spectral resolution of ~3.1 nm and bandwidth of ~50 nm, which results in the limit of detection of 0.042 RIU. The MRR output spectrum with air- and water-claddings are measured and reconstructed from the MRR-SHFTS integrated device experimentally to validate the wavelength shifting measurement.

## I. INTRODUCTION

As the needs of the accurate and fast point-of-care portable bio-detection systems are growing rapidly for the clinical and health-monitoring applications, various of micro- and nano-scale optical biosensors have intrigued a significant attention as compact, highly selective, and sensitive real-time biosensor platforms. Among them, micro-ring-resonator (MRR) based optical sensors have been demonstrated in numerous applications due to their advantages of high sensitivity and small footprint compared to the other photonic biosensor platforms [1-3]. The general sensing principle of MRR bio sensing is based on the detection of the resonance wavelength shift ($\Delta\lambda$). Based on the evanescent field sensing mechanism, the resonance peaks of the MRR are shifted when the refractive index of the functionalization layer is changed due to the binding between the target bio-molecules and specific immobilized bioreceptors on the MRR surface; this way the sensing specificity is achieved due to the selective binding. Several research have presented highly-sensitive MRR bio-sensing applications in near-infrared wavelength with advanced structures including sub-wavelength grating based waveguides [4,5], and cascaded resonators utilizing Vernier effects [6-8] to maximize the sensing sensitivity; but utilizing external optical spectrum analyzer (OSA) was inevitable to read the spectrum and detect $\Delta\lambda$ in any case, which makes the overall system bulky and expensive eventually. To miniaturize the whole system into a chip for the actual lab-on-a-chip biosensor, the integration of MRR and on-chip spectrometer benefit from the photonic integrated circuits is highly demanded; by monolithically integrate the MRR with the on-chip spectrometer, it is able to remove the fiber coupling alignment or any other moving components between the resonator and spectrometer, which makes the device safe from the environmental perturbations, in turn providing more reliable, robust, and low-cost advantages.

Several on-chip spectrometers have been demonstrated including the dispersive optics-based spectrometers [9-13], Fourier transform spectrometers (FTS) [14-19] and active tuning (electro/thermo-optic) based spectrometers [20-22]; for the point of care (PoC) portable biosensing applications which can read the spectrum in real-time, high-speed on-chip spectrometer based on the PICs is necessary, which makes the FTS an attractive solution that measures the spectrum with the interference of light instead of dispersion in a single capture, offering the advantages of larger SNR and faster date collection speed. Among the on-chip FTS schemes, the spatial heterodyne-FTS (SHFTS) consists of an array of unbalanced Mach-Zehnder interferometers (MZIs) with linearly increasing optical path differences (OPDs). Based on the interferometric schemes, the phase changes from each MZI are converted into the intensity change of each MZI output powers, which can be measured by a photodetector array in a single capture without any active-modulator or moving components. Consequently, the input spectrum can be reconstructed by Fourier-transforming the output power interferogram measured from the MZI array [16]. However, the spectral resolution and bandwidth of SHFTS are highly limited by the number of MZIs and maximum OPD, and considering the on-chip integration with MRR for the PoC biosensing applications, the SHFTS have to be carefully designed to fully reconstruct the spectrum covering the free spectral range (FSR) of the MRR and resonance wavelength shift.

The silicon-on-insulator (SOI) with the buried $SiO_2$ cladding is the most matured and attractive material platforms for the photonic biosensors because of the possibilities of scalable-mass production and compatibilities with a covalently attached functionalization layer coating [23]. Typical operation wavelengths of SOI-based photonic biosensors are within the optical C-band ($\lambda$=1550 nm), because the SOI is highly transparent, and many low-cost light sources are already available in the market at these wavelengths.

In this work, we designed the MRR biosensors integrated with the on-chip SHFTS device using the CMOS compatible SOI wafer, and experimentally demonstrated the resonance shift retrieval results from the fabricated SHFTS-MRR device in C-band.

## II. DEVICE DESIGN

The basic device configuration and working principle are demonstrated in Figure 1. The MRR is a biosensing device which configures the resonance peaks from the drop-port output, and the evanescent field is altered when the molecular binding takes place between the immobilized bioreceptors and target analytes in the sample which changes the resonance condition leading to a resonance wavelength shift (Δλ). The drop-port of MRR is directly connected to the SHFTS, that is composed of an array of unbalanced MZIs with linearly increasing optical path delays ($\Delta L_i$). The phase change from each MZI is converted into an intensity change based on interferometric schemes, constituting the interferogram of the output powers. Output powers from each MZIs can be measured by the integrated photodetector (PD) array [24], and the input spectrum from the MRR drop-port is reconstructed through the discrete Fourier transform equation (DFT) using the computation program, such as MATLAB.

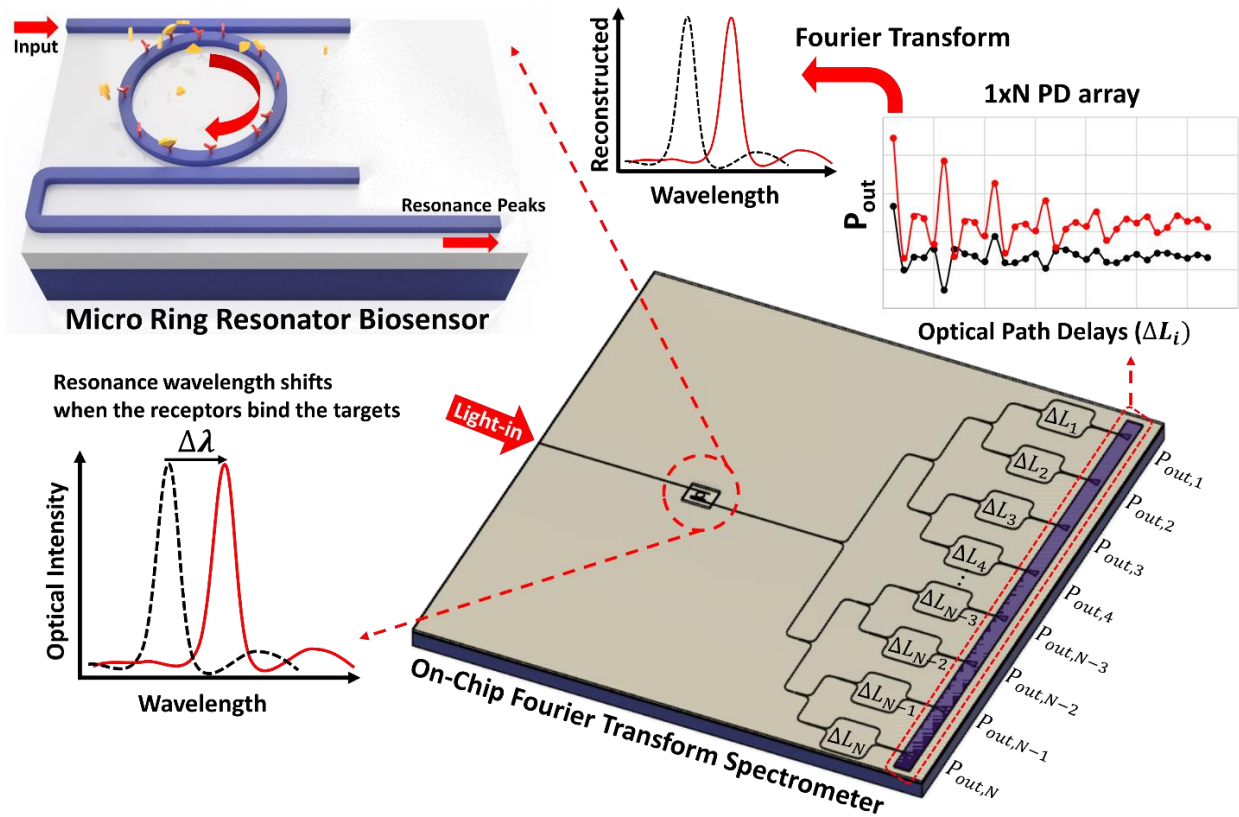

**Figure 1.** Schematic illustration of the operation principle of the on-chip spectrometer integrated optical biosensor platform; PD: photodetector.

The center operation wavelength is 1550 nm, and the layer thicknesses of the SOI wafer and the strip waveguide are designed for guiding the fundamental transverse-electric (TE) mode based on the previous research [4,5,25]; the width and thickness of the silicon strip waveguide are 500 nm and 220 nm, respectively [4,5], and the thickness of SiO2 bottom cladding is ~3 $\mu$m which is optimized for the grating coupler design for the fundamental TE mode based on the previous focusing sub-wavelength grating coupler (SWGC) structures [25]. Based on the single mode strip waveguide structure, we designed and fabricated the MRR and SHFTS devices and experimentally tested the device performances.

**2. 1. Silicon Micro Ring Resonator Design**

As mentioned in the introduction, the basic sensing mechanism of MRR biosensor is based on the evanescent field sensing; the resonance peaks shift when the refractive index of the top cladding (functionalized layer) changes because of the bio receptor-target molecules binding. Generally, the bulk sensing sensitivity (S) of the MRR biosensor is defined as [2]:

$$S = \frac{\Delta\lambda}{\Delta n_{clad}} \quad \text{(eq. 1)}$$

where $\Delta\lambda$ is the resonance wavelength shift, and $\Delta n_{clad}$ is the change of the cladding refractive index. Also, the limit-of-detection (LOD) of the sensor depends on the minimum resolvable wavelength shift, that is determined by the measurement setup [2]:

$$LOD = \frac{\Delta\lambda_{min}}{S} \quad \text{(eq. 2)}$$

Here, $\Delta\lambda_{min}$ is the minimum detectable resonance wavelength shift, which is the optical spectrum analyzer's measurement resolution ($\delta\lambda$). To maximize the sensing sensitivity, various of the waveguide structures and polarization effects have been studied, including the strip waveguide, slot waveguide [26], and sub-wavelength grating waveguide [4,5] with TE or TM (transverse-magnetic) mode. Essentially, the more light is interacting with the cladding (analytes), the higher is the sensing sensitivity, but the optical losses are increasing at the same time, which hinder to achieve high-quality resonance peaks [2]. Hence, a good compromise between the MRR sensing sensitivity and the optical loss of the structure have to be considered [27]. However, regardless of the MRR sensitivity improving strategies, this work focuses on the integration of MRR biosensor devices and SHFTS for the monolithic integration of the sensor and the spectrometer for the sensing-reading system integration, and we designed the basic symmetric add-drop ring resonator device based on the strip waveguide with fundamental TE polarization.

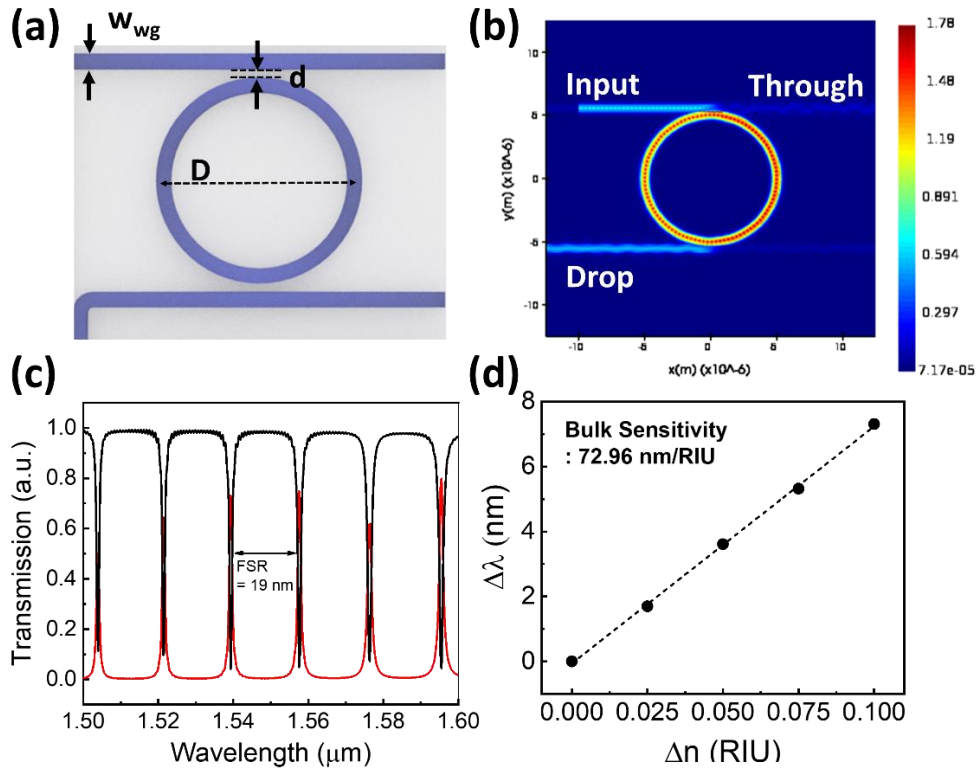

**Figure 2.** (a) MRR biosensor design. (b) Electric-field simulation at resonance peak. (c) Transmission spectrum from the through port and drop port; black-line: through port, red-line: drop port. (d) Resonance wavelength shift based on the refractive index change of the top-cladding.

The design parameters include the gap-distance (d) between the bus-waveguide and ring-waveguide to ensure the critical coupling based on the couple mode theory [28], and the diameter of the ring resonator (D) is designed to get the large enough FSR to make it compatible with the SHFTS spectral bandwidth coverage. The FSR, which is the wavelength range between two resonances can be calculated as follow [1]:

$$FSR = \frac{\lambda^2}{n_g L} \quad (eq.\ 3)$$

where $n_g$ is the group index of the waveguide, and the L is the round-trip length of the ring waveguide. Another important optical characteristic of the MRR is the quality factor (Q), which is a measure of the sharpness of the resonance peak that is defined as follow [1]:

$$Q = \frac{\lambda_{res}}{FWHM} \quad (eq.\ 4)$$

where $\lambda_{res}$ is the resonance wavelengths, and FWHM is the full width at half maximum of the resonance spectrum.

The schematic illustration of the symmetric add-drop MRR device is shown in Figure 2(a), and the device parameters are optimized by the 3-dimensional (3D) finite-difference time-domain (FDTD) simulation, as d=70 nm and D=10 $\mu$m. The top view of the E-field simulation result at the resonance peak is shown in Figure 2(b), showing that the incident field at the resonance wavelength is transmitted to the drop port. The transmission spectrum of the drop- and through-ports are monitored by the 3D FDTD simulation (Figure 2(c)), and the optical characteristics are measured as FSR=19 nm, and Q≅4000 from optimized structure. Then, the refractive index of the top cladding is changed ($\Delta n_{clad}$) and the resonance wavelength shift ($\Delta\lambda$) is measured to calculate the bulk-sensing sensitivity of the biosensor that is presented in Figure 2(d). The bulk-sensing sensitivity is calculated as S=72.96 nm/RIU.

## 2. 2. Silicon SHFTS for Biosensing Applications

The theory and principle of the standard SHFTS have been demonstrated by Florjańczyk et al. [16]. As briefly mentioned in the introduction, SHFTS consists of an array of unbalanced MZI with linearly increasing OPD with constant increment across the array configuring spatial interferogram. For a given single-input source, the phase change from each MZI is converted into an intensity change based on interferometric schemes. The input spectrum can be retrieved through the discrete Fourier transform (DFT) which can be written as [16]:

$$p^{in}(\bar{\sigma}) = \frac{\Delta x}{N} P^{in} + 2 \frac{\Delta x}{N} \sum_{i=1}^{N} F(x_i) \cos 2\pi \bar{\sigma} x_i, \text{ where } F(x_i) = \frac{1}{B_s}(2P_i^{out} - A_s P^{in}) \quad (eq.\ 5)$$

Here, $P^{in}$ is the input power, N is the number of MZIs, $\sigma$ is the wavenumber, and $\bar{\sigma} = \sigma - \sigma_{min}$ is the shifted wavenumber, where $\sigma_{min}$ represents the minimum wavenumber at the Littrow condition; at the Littrow condition, the phase delays in different MZIs are integer multiplies of $2\pi$, so the output powers of each MZI ($P_i^{out}$) are constant. $\Delta x$ is the maximum interferometric delay, that is $\Delta x = n_{eff} \Delta L_{max}$, where $n_{eff}$ is the effective index of the strip waveguide and $\Delta L_{max}$ is the maximum path delay of the most unbalanced MZI. The spatial interferogram $F(x_i)$ is discretized at N equally spaced OPD values $x_i$ ($0 \leq x_i \leq \Delta x$) which is defined as $x_i = n_{eff} \Delta L_i$, where $\Delta L_i$ is the path length difference of the $i$ th unbalanced MZI. The input power $P^{in}$ is constant for all the MZIs, and $P_i^{out}$ represents the output power of the $i$ th MZI with the coupling and loss coefficients of the MZI components $A_s$ and $B_s$. As the wavenumber of monochromatic input $\sigma$ changes from the Littrow wavenumber, $P_i^{out}$ distribution becomes periodic, and different wavenumbers create different periodic patterns. Subsequently, a polychromatic input signal, which can be considered as a superposition of monochromatic constituents, creates a corresponding spatial interferogram pattern formed by a superposition of the respective periodic $P_i^{out}$ fringes from monochromatic input. The resolution of spectrometers, represented by the wavenumber resolution $\delta \sigma$ is determined by the maximum interferometric delay $\Delta x$. To resolve two monochromatic signals separated by $\delta \sigma$ at the most unblanced MZI with OPD of $\Delta x$, the phase change from respective interferograms of $\sigma$ and $\sigma + \delta \sigma$ should be differ by one fringe ($2\pi$), that is $\Delta \varphi = 2\pi(\sigma + \delta \sigma)\Delta x - 2\pi \sigma \Delta x = 2\pi$, in turn, [16]

$$\delta\sigma\Delta x = \delta\sigma\Delta L_{max}n_{eff} = 1 \qquad (eq.\ 6)$$

Thus, the maximum path delay of the MZI array ($\Delta L_{max}$) can be designated as follows:

$$\Delta L_{max} = \frac{1}{\delta\sigma \cdot n_{eff}}, where\ \delta\sigma = \frac{1}{\lambda_o} - \frac{1}{\lambda_o + \delta\lambda} \qquad (eq.\ 7)$$

where $\lambda_o$ is the center wavelength and $\delta\lambda$ is the wavelength resolution. The number of MZIs in the array (N) is equivalent to the discrete sampling points in the spatial interferogram. Based on the Nyquist-Shannon sampling theorem, the minimum sampling points ($N_{min}$) required to fully reconstruct the input spectrum ($p^{in}(\bar{\sigma})$) within the band-limit range is determined as follows:

$$N_{min} = 2\Delta x \Delta\sigma = 2\frac{\Delta\sigma}{\delta\sigma} \qquad (eq.\ 8)$$

where $\Delta\sigma$ is the bandwidth of the spectrometer.

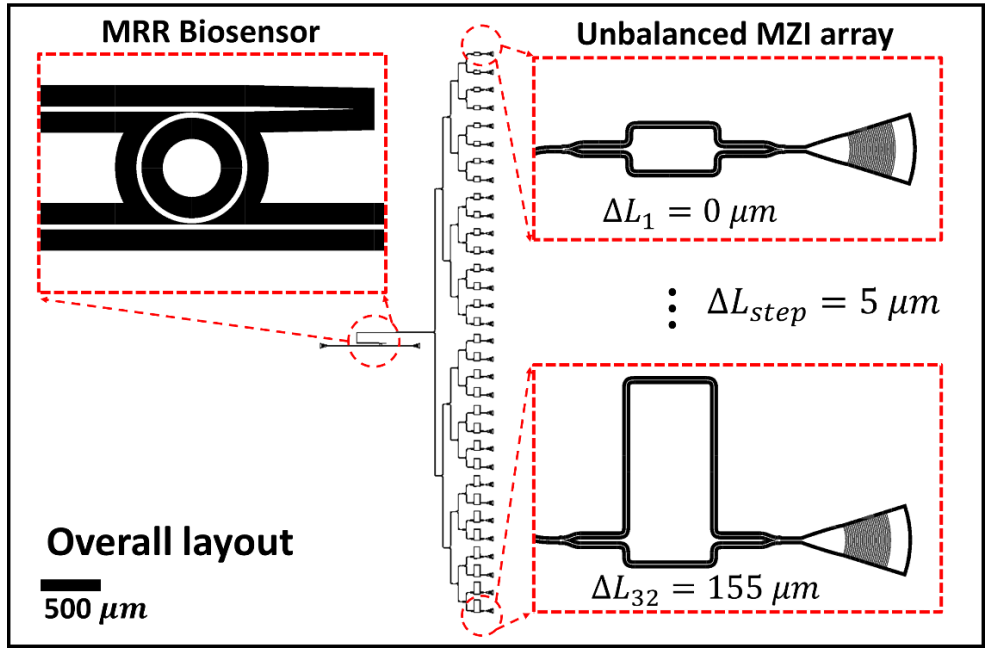

**Figure 3.** Overall layout of the MRR-SHFTS integrated biosensor device.

For the biosensing application integrated with MRR, the spectrometer should be able to read and distinguish the separated resonance peaks by the FSR completely. Accordingly, the silicon SHFTS is designed to reconstruct the resonance peaks from the optimized MRR shown in Figure 2, in turn to have at least ~40 nm of bandwidth to retrieve at least two resonance peaks without aliasing errors, and the resolution is designed to be smaller than ~5 nm to make sure to resolve each peak separated by FSR clearly. By using the (eq. 7) and (eq. 8), the silicon SHFTS is designed to have 32-MZIs (N=32) with the maximum path length delay (ΔL) of 155 μm, which gives us the spectral bandwidth of 50 nm, and the resolution of 3.1 nm, respectively. Subsequently, based on the sensing sensitivity of the MRR biosensor presented in chapter 2.1, the LOD of the MRR-SHFTS integrated sensor is calculated as LOD=0.042 RIU by (eq. 2), which is the minimum detectable refractive index change of the top cladding ($\Delta n_{clad}$). At this point, compared to the conventional OSA which typically provides the wavelength resolution ±0.02 nm, the resolution of the SHFTS has to be improved to enhance the LOD of the biosensor, however the resolution of the standard SHFTS is highly limited by the number of MZIs (N); in order to achieve ~30 pm resolution while maintaing the bandwidth of 50 nm, approximately 3200-MZIs are requird, which make the size of the device significantly. Instead of having an outrageous number of MZIs, several studies have proposed utilizing the active photonic components introducing the thermal/electrical optical phase delay [20-22]; among them, lithium niobate (LN) based electro-optics modulator is one of the promising candidates to improve the resolution and bandwidth of the spectrometer in terms of high-speed which is beneficial for the biosensing application [22].

Our prototype device focuses on the proof-of-concept of the sensor-spectrometer integration, and we used the SWGC [25] for coupling the input light source and collecting the powers from the MZI outputs, and the final device layout is presented in Figure 3, which shows the fully integrated MRR biosensors and SHFTS device. Based on the optimized designs, we have fabricated and experimentally measured the optical characteristics and reconstructed the MRR resonance peaks from the SHFTS.

## III. DEVICE FABRICATION

Every component of the MRR-SHFTS is a through-etched structure, so it was able to pattern and etch the whole device in one lithography and single silicon-etching step. The SOI wafers have a 220 nm thick Si on a 3 μm thick $SiO_2$ bottom cladding on a Si substrate. Then, ~400 nm thick Ebeam resist (ZEP-520A) is deposited by spin-coating. The patterning is done by JEOL E-beam (JBX-6000FS) lithography tool, followed by developing in n-Amyl acetate for 2 min, and rinsing in isopropyl alcohol (IPA). Following this, the pattern is transferred to Si layer by inductively coupled plasma etching (ICP-RIE). Finally, the remaining resist and polymers are cleaned using removal PG and followed with cycles of Acetone/IPA post process treatment. The pictures of the fabricated MRR-SHFTS device including the overall layout and the SEM images of SWGC and MRR are shown in Figure 4.

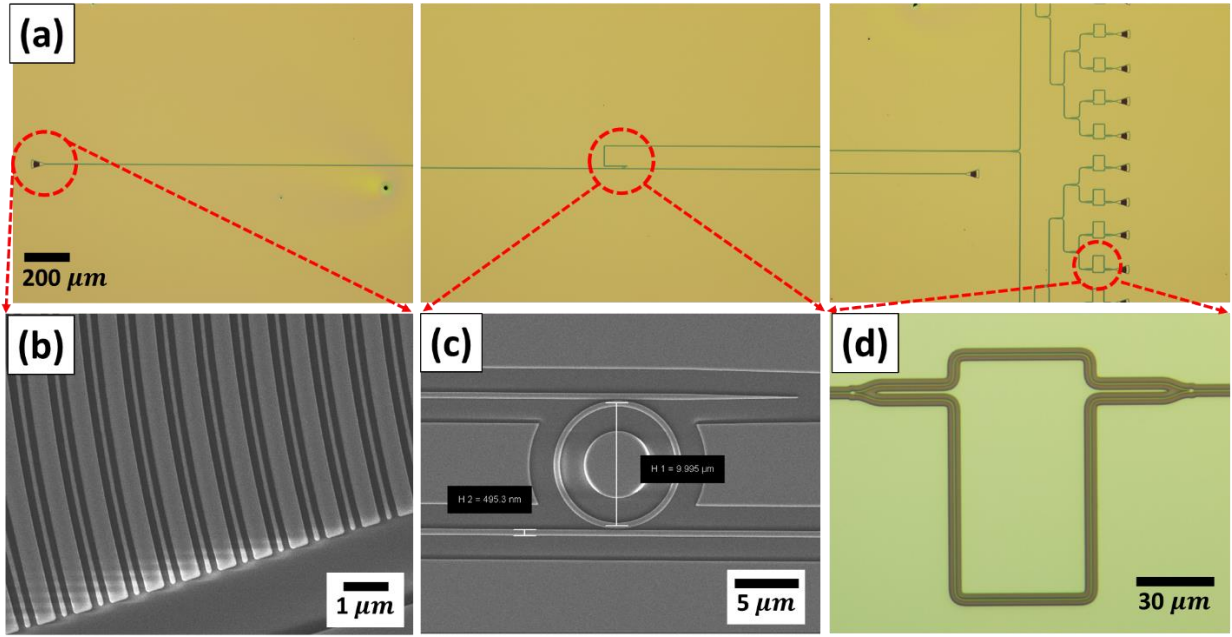

**Figure 4.** Fabricate device images. (a) Microscope image of the overall device. (b) SEM image of the input SWGC and (c) MRR. (d) Microscope image of one of the unbalanced MZI.

## IV. DEVICE CHARACTERIZATION

## 4. 1. MRR Transmission Spectrum Measurement

First of all, the transmission spectrum of the fabricated MRR device shown in Figure 4(c) is measured using the OSA directly. Figure 5(a) shows the measurement setup. The NIR broadband light source (ASE-FL7001P) emits the light with the wavelength from 1.52 $\mu m$ to 1.64 $\mu m$ and coupled to the input SWGC through the single-mode fiber (SMF). Then, the MRR output spectrum is collected from the output through- and drop-ports SWGCs through another SMF and measured by the OSA. Figure 5(b) shows the MRR output spectrum measurement results, including the through- and drop-ports; the envelope shape of the through-port signal is determined by the SWGC coupling. As a result, we measured the FSR=19 nm and Q~4000 from the experimental results.

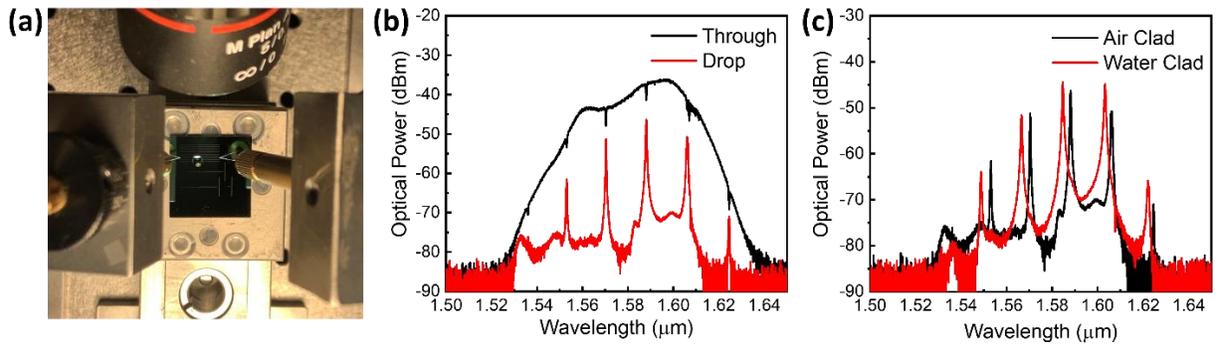

**Figure 5.** MRR measurement results. (a) Measurement setup picture; water-droplet is placed on top of the MRR. (b) OSA measurement results from the through- and drop-port with air cladding. (c) OSA measurement results from the drop-port with air and water cladding.

Next, we tested the resonance wavelength shift due to refractive index changes from the air- and water-cladding on top of the MRR device experimentally. Figure 5(c) shows the measured spectrum from the drop port with air- and water- cladding with $\Delta\lambda = 16\ nm$.

## 4. 2. Tunable Laser Source Spectrum Reconstruction From SHFTS

Secondly, the fabricated SHFTS device was tested with the tunable laser source to validate the spectrum reconstruction performance. The tunable laser source (CoBrite DX4 from ID Photonics

GmbH) provides the monochromatic signal with tunable wavelength range from 1530 nm to 1567 nm, with the output power of ~250 $\mu$W measured from the SMF. For a given single input, each monochromatic wavelength creates a corresponding spatial interferogram fringe output pattern from 32-unblanced MZIs with linearly increasing OPD. We measured the output powers from each MZI with the wavelength of 1550 nm, 1555 nm and 1560 nm respectively, which are shown in Figure S1. Based on the measured interferograms, we reconstructed the optical spectrum of the each monochromatic signal by the MATLAB code based on DFT equation (eq. 5), and the results are shown in Figure 6. To validate the spectrum retrieval accuracy, the FTS retrieved results (black-dashs) were compared with the direct OSA measurement results (red-lines).

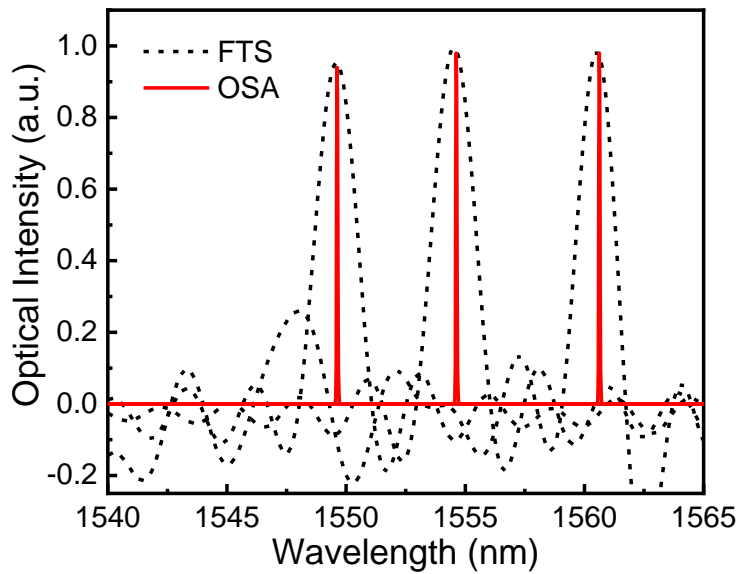

**Figure 6**. Reconstructed spectrum of the tunable laser source at 1550 nm, 1555 nm, and 1560 nm; Red-line: OSA measurements, Black-dash: FTS reconstructed.

The wavelength positions of FTS reconstructed spectra are well matched with the direct OSA measurement results but the resolution of SHFTS is limited as ~3.1 nm, which result in broader shape in reconstructed spectrum, and the discrepancies and ripples from the reconstructed spectrum are mainly due to the optical phase errors induced from the etching surface and sidewall roughness.

### 4. 3. MRR-SHFTS Spectrum Reconstruction Results

Finally, by integrating the MRR biosensor and SHFTS which are demonstrated in earlier chapters, we reconstructed the drop-port output signals of MRR by SHFTS and read the resonance wavelength shift with different claddings. The resonance peaks from the MRR drop-ports shown in Figure 5(c) are the polychromatic input signals of SHFTS, which can be considered as a superposition of monochromatic constituents, creates a corresponding spatial interferogram pattern formed by a superposition of the respective periodic $P_i^{out}$ fringes from monochromatic input. The measured output powers with water- and air-cladding are shown in Figure S2. Then, following the same DFT equation (eq. 5), the spectrum from the MRR with different claddings are reconstructed, and compared with the direct MRR spectrum measurement results by OSA as shown in Figure 7.

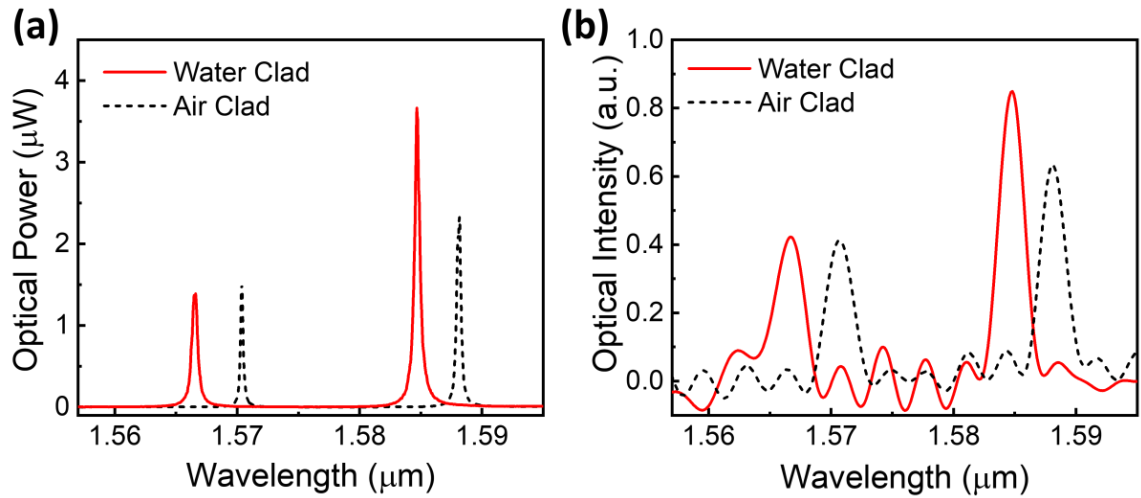

**Figure 7.** MRR-SHFTS measurement result. (a) OSA measured spectrum of the MRR drop-port. (b) FTS reconstructed spectrum.

Figure 7(a) shows the MRR drop-port spectrum measurement results by OSA with different claddings, and Figure 7(b) shows the SHFTS-MRR reconstructed spectrum. As the FTS reconstucted spectrum shows well-matching results with the direct OSA measurement results, we were able to validate that the MRR-SHFTS integrated device can succesfully substitute the external OSA for the lab-on-a-chip biosensor applications. However, the resolution of the current SHFTS configuration is limited as ~3 nm, which limits the LOD=0.042 RIU. To enhance the limit of detection, both the sensing sensitivity of MRR and the resolution of the SHFTS have to be improved as discussed in the earlier chapter 2.1 and 2.2. Moreover, our prototype device can be further improved by integrating the photodetector array [24] as shown in Figure 1, and the on-chip light sources for the NIR silicon photonics have been researched thoroughly to provide the SOI compatible cost-effective solutions, and several hybrid on-chip Si laser schemes have been introduced previously [29-31]. Thereby, in combination with these on-chip light sources and detector devices, we expect the fully-packaged monolithic integrated photonic biosensors will be available by removing the fiber alignment or any other additional alignment processes, which makes the device safe from the environmental perturbations, in turn providing sensitive, reliable, robust and low-cost sensor platforms.

## V. CONCLUSION

We experimentally demonstrated the integration of MRR biosensor and on-chip SHFTS device for the lab-on-a-chip biosensor platform. The symmetric add-drop MRR is designed to have FSR=19 nm, Q~4000, and S=72.9 nm/RIU and the SHFTS is designed to have the resolution and bandwidth coverage of ~3nm and ~50nm respectively. The resonance peaks from the MRR drop-port are reconstructed from the SHFTS, and the resonance wavelength shift due to the refractive index change is retrieved from the reconstructed spectrum with the limit of detection of 0.042 RIU.

Our proof-of-concept experiment demonstrates the on-chip biosensing platform without using the external OSA to read the wavelength data and paves the way for a monolithically integrated lab-on-a-chip optical biosensor with the light source/sensor/detector integration.

**Acknowledgements**

This research was supported by NSF Award #1932753, and the authors wish to thank the anonymous reviewers for their valuable comments and suggestions.

**Author Contributions**

**Kyoung Min Yoo**: Conceptualization (lead), Data Curation (lead), Formal Analysis (lead), Methodology (lead), Project Administration (lead), Supervision (equal), Validation (equal), Visualization (lead), Writing-Original Draft Preparation (lead), Writing-Review and Editing (lead). **May Hlaing**: Validation (supporting). **Sourabh Jain**: Validation (supporting). **James Fan**: Validation (supporting). **Yue An**: Validation (supporting). **Ray T. Chen**: Conceptualization (supporting), Funding Acquisition (lead), Resources (lead), Supervision (equal), Validation (equal), Writing-Review and Editing (supporting).

**Conflict of Interest**

The authors declare no competing interests.

**Data Availability**

The data that support the findings of this study are available from the corresponding author upon reasonable request.

**Supporting Information Available:**

See the supplementary material for the measured output powers from the SHFTS.

# Supporting Information

# Lab-on-a-Chip Optical Biosensor Platform: Micro Ring Resonator Integrated with Near-Infrared Fourier Transform Spectrometer

KYOUNG MIN YOO[1,3], MAY HLAING[2], SOURABH JAIN[1], JAMES FAN[1], YUE AN[1], AND RAY T CHEN[1,2,4]

[1]Department of Electrical and Computer Engineering, The University of Texas at Austin, 10100 Burnet Rd. Austin, TX, 78758, USA.
[2]Omega Optics Inc., 8500 Shoal Creek Blvd., Bldg. 4, Suite 200, Austin, TX, 78757, USA.
[3]yoo_eb@utexas.edu
[4]chenrt@austin.utexas.edu

This supplementary material includes the spatial heterodyne Fourier transform spectrometer (SHFTS) output power measurement data from 32-MZIs; Figure S1 shows the SHFTS output power interferogram using the tunable laser source (CoBrite DX4 from ID Photonics GmbH), which provides the monochromatic signals with the $\lambda = 1550\ nm, 1555\ nm, and\ 1560\ nm$.

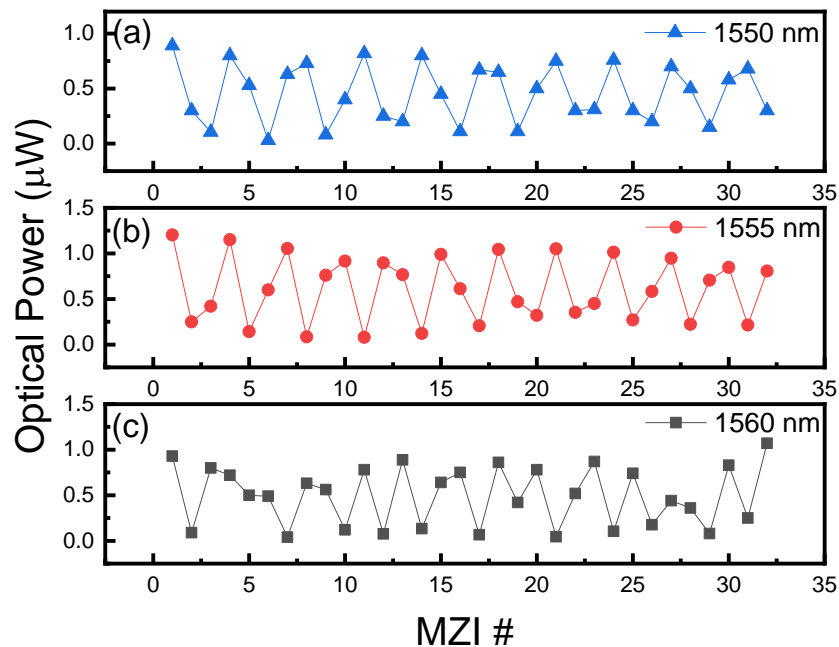

Figure S1: SHFTS output power measurement results using the tunable laser source at the wavelength of (a) 1550nm, (b) 1555 nm, and (c) 1560 nm.

Figure S2 shows the output interferogram from the MRR-SHFTS with two different top claddings using the NIR broadband light source (ASE-FL7001P), which emits the light with the wavelength from 1.52 $\mu m$ to 1.64 $\mu m$.

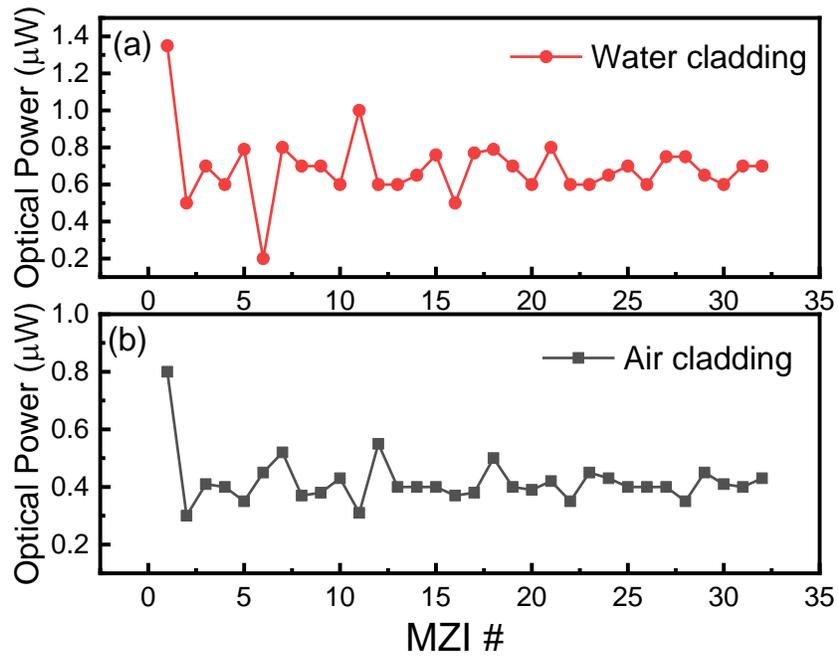

Figure S2: MRR-SHFTS output power measurement results with (a) water- and (b) air-claddings.